\documentclass[12pt]{article}
\usepackage{a4wide}
\usepackage{graphicx}

\begin{document}
\begin{flushright}
MZ-TH/11-45\\
December 2011
\end{flushright}

\begin{center}
{\Large\bf Top quark polarization at a\\[7pt]
polarized linear $e^{+}e^{-}$ collider}\\[1truecm]
{\large S.~Groote$^{1,2}$ and J.G.~K\"orner$^1$}\\[.5truecm]
$^1$ Institut f\"ur Physik, Johannes-Gutenberg-Universit\"at,
  Mainz, Germany\\[.5truecm]
$^2$ F\"u\"usika Instituut, Tartu \"Ulikool, Tartu, Estonia
\vspace{1truecm}
\end{center}

\begin{abstract}
We discuss the polarization of top quarks produced at a polarized 
linear $e^{+}e^{-}$--collider. Close-to-maximal values of the top quark 
polarization can be achieved with longitudinal beam polarizations 
$(h_{-}\sim-0.80,\,h_{+}\sim+0.625)$ 
or $(h_{-}\sim +0.80,\,h_{+}\sim -0.625)$ at intermediate beam energies.
The option $(h_{-}\sim-0.80,\,h_{+}\sim+0.625)$ has to be preferred since
this choice is quite stable against variations of the beam polarization.
All our quantitative results have been obtained at NLO \,QCD.
\end{abstract}

\newpage

\section{Introductory remarks}
It is well--known that the top quark keeps its polarization acquired in 
production when it decays since 
$\tau_{\rm hadronization} \gg \tau_{\rm decay}$.
One can test the Standard Model (SM) and/or non-SM couplings through 
polarization measurements involving top quark decay 
(mostly  $t \to b + W^{+}$). New observables involving top quark
polarization can be defined such as 
$<\vec{P}\cdot \vec{p}>$ (see e.g. 
\cite{hep-ph/9710225,hep-ph/9811482,hep-ph/0101322,hep-ph/0602026,
arXiv:1005.5382,arXiv:1010.2402}).
It is clear that the analyzing power of such observables is largest for
large values of the polarization of the top quark $|\vec{P}|$. This calls for
large polarization values.
It is, nevertheless, desirable to have a control sample with small polarization
of the top quark. In this talk we report on the results 
of investigations in~\cite{arXiv:1012.4600} whose aim was to find maximal 
and minimal values of top quark polarization at a linear 
$e^{+}e^{-}$--collider  by tuning the longitudinal beam 
polarization~\cite{arXiv:1012.4600}. At the same time one wants to keep the
top quark pair production cross--section large. It is a lucky coincidence that
all these goals can be realized at the same time.

\begin{figure}[ht]
\begin{center}
\hspace{-1.0cm}
\includegraphics[scale=0.8]{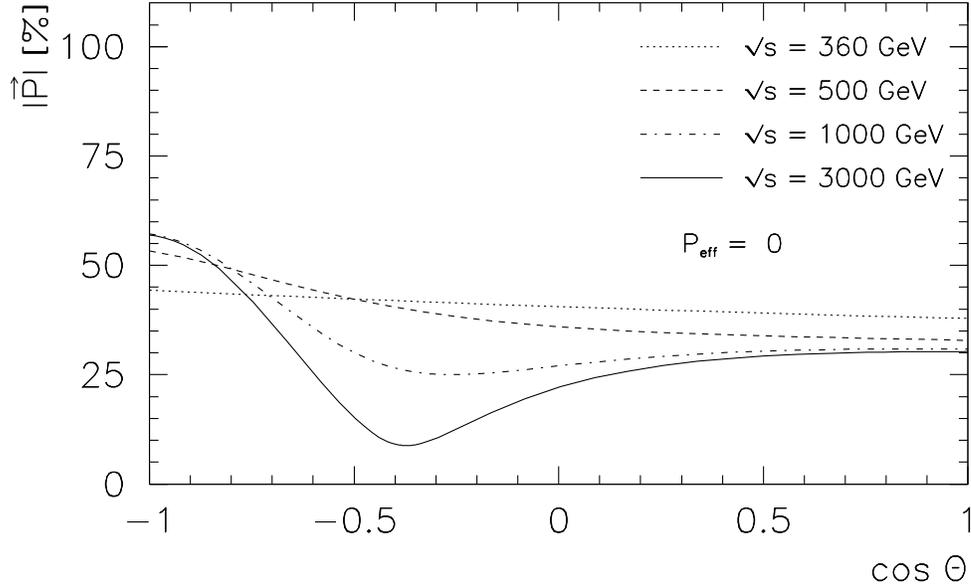} 
\end{center}
\caption{\label{fig:zeropol}Total NLO top quark polarization for zero beam 
polarization
}
\end{figure}

Let us remind the reader that the top quark is polarized even for zero beam 
polarization through vector-axial vector interference effects
 $\sim v_{e}a_{e},\,v_{e}a_{f},\,v_{f}a_{e},\,v_{f}a_{f}$, where 
\begin{eqnarray}
\label{ewcouplings}
v_{e},a_{e}\quad&:& \mbox{electron current coupling}\\
v_{f},a_{f}\quad&:& \mbox{top quark current coupling}
\end{eqnarray}
In Fig.~\ref{fig:zeropol} we plot the $\cos\theta$ dependence of the zero
beam polarization top 
quark polarization for different characteristic energies at 
$\sqrt{s}=360$ GeV (close to threshold), $\sqrt{s}=500$ GeV (ILC phase 1),
$\sqrt{s}=1000$ GeV (ILC phase 2) and $\sqrt{s}=3000$ GeV (CLIC). We mention
that the planning
of the ILC has reached a stage where the Technical Design Report (TDR) will
be submitted in 2012.
%%%%%%%%%%%%%%%%%%%%%%%%%%%%%%%%%%%%%%%%%%%%%%%%%%%%%%%%%%%%%%%%%%%%%%%%%%%%%
%%%%%%%%%%%%%%%%%%%%%%%%%%%%%%%%%%%%%%%%%%%%%%%%%%%%%%%%%%%%%%%%%%%%%%%%%%%%%

\section{Top quark polarization at threshold\\ and in the high energy limit}
The polarization of the top quark depends on the c.m. energy $\sqrt{s}$,
the scattering angle $\cos\theta$, the electroweak coupling coefficients 
$g_{ij}$ and the effective beam polarization $P_{\rm eff}$, i.e. one has
\begin{equation}
\label{vecpol}
\vec{P}=\vec{P}\,(\sqrt{s},\cos\theta, g_{ij},P_{\rm eff})\,,
\end{equation}
where the effective beam polarization appearing in
(\ref{vecpol}) is given by
\begin{equation}
\label{peff}
P_{\rm eff}=\frac{h_- - h_+}{1 - h_- h_+}\,.
\end{equation}
and where $h_-$ and $h_+$ are the longitudinal polarization of the
electron and positron beams $(-1<h_{\pm}<1)$, respectively.

For general energies the functional dependence Eq.(\ref{vecpol}) is not 
simple. Even if the electroweak couplings $g_{ij}$ are fixed one remains with 
a three--dimensional parameter space $(\sqrt{s},\cos\theta,P_{\rm eff})$. 
However, the polarization formula 
considerable simplifies at nominal threshold $\sqrt{s}=2m$ and in the high 
energy limit $\sqrt{s} \to \infty$.

At threshold and at the Born term level one has
\begin{equation}
\label{thresh}
\vec{P}_{\rm thresh}=\frac{P_{\rm eff}- A_{LR}}{1-P_{\rm eff}A_{LR}}\,\,\,\hat{n}_{e^-}\,,
\end{equation}
where $A_{LR}$ is the left-right beam polarization asymmetry
$(\sigma_{LR}-\sigma_{RL})/(\sigma_{LR}+\sigma_{RL})$ at threshold
and $\hat{n}_{e^-}$ is a unit vector pointing into
the direction of the electron momentum. We use a notation where
$\sigma(LR/RL)=\sigma(h_{-}=\mp 1;h_{+}=\pm 1)$. In terms of the electroweak 
coupling parameters $g_{ij}$, the nominal polarization 
asymmetry at threshold $\sqrt{s}=2m_t$ is given by 
$A_{LR}=-(g_{41}+g_{42})/(g_{11}+g_{12})=0.409$. Eq.(\ref{thresh}) shows
that, at threshold and at the Born term level, the polarization $\vec{P}$ 
is parallel to the beam axis irrespective of the scattering angle and has 
maximal values $|\vec{P}|=1$ for $P_{\rm eff}=\pm1$. Zero polarization is
achieved for $P_{\rm eff}=A_{LR}=0.409$.

In the high energy limit the polarization of the top quark is purely 
longitudinal, i.e. the polarization points into the direction of the top quark.
At the Born term level one finds 
$\vec{P}(\cos\theta)=P^{(\ell)}(\cos\theta)\cdot\hat{p_t}$ with
\begin{eqnarray}
\label{helimit}
P^{(\ell)}(\cos\theta)&=& \nonumber\\&&\hspace*{-2.3cm}
\frac{(g_{14}+g_{41}+P_{\rm eff}(g_{11}+g_{44}))(1+\cos\theta)^2
  +(g_{14}-g_{41}-P_{\rm eff}(g_{11}-g_{44}))(1-\cos\theta)^2}
  {(g_{11}+g_{44}+P_{\rm eff}(g_{14}+g_{41}))(1+\cos\theta)^2
  +(g_{11}-g_{44}-P_{\rm eff}(g_{14}-g_{41}))(1-\cos\theta)^2}\nonumber\\
\end{eqnarray}
In the same limit, the electroweak coupling coefficients appearing in 
(\ref{helimit}) take the numerical values $g_{11}=0.601$,
$g_{14}=-0.131$, $g_{41}=-0.201$ and $g_{44}=0.483$.
%$g_{12}=0.352$ and $g_{42}=-0.164$.

It is quite evident that the two limiting cases have quite a different
characteristics and different functional behaviour. The question is whether the
two limiting cases can be taken as guiding principles for intermediate
energies and for which. The answer is yes and no, or sometimes. 

Take, for example, the differential $\cos\theta$
rate which is flat at threshold and shows a strong forward peak in the high
energy limit with very little dependence on $P_{\rm eff}$. This can be seen 
by substituting the numerical high energy values of the gauge couplings 
$g_{ij}$ in the denominator of Eq.(\ref{helimit}). One finds 
\begin{equation}
\label{forrate}
\frac{d\Gamma}{d\cos\theta}(s \to \infty)\,\propto\,(1.084 -0.332 P_{\rm eff})
(1+\cos\theta)^{2}
+(0.118 -0.007P_{\rm eff})
(1-\cos\theta)^{2}\,. 
\end{equation}
More detailed calculations show that the strong forward dominance of the rate
sets in rather fast above threshold~\cite{arXiv:1012.4600}. This is quite 
welcome since the forward region is favoured from the polarization point 
of view.

As another example take the vanishing of the polarization which, at threshold,
occurs at $P_{\rm eff}=0.409$. In the high energy limit, and in the forward
region where the numerator part of (\ref{helimit}) proportional to
$(1+\cos\theta)^2$ dominates, one finds a polarization zero at
$P_{\rm eff}=(g_{14}+g_{41})/(g_{11}+g_{44})=0.306$. The two values of 
$P_{\rm eff}$ do not differ much from another.
\section{Effective beam polarization}
Let us briefly recall how the effective beam polarization $P_{\rm eff}$
defined in Eq.(\ref{peff}) enters the description of polarized beam effects. 
Consider the rates
$\sigma_{LR}$ and $\sigma_{RL}$ for $100\%$ longitudinally polarized
beams. The rate $\sigma(beampol)$ for partially polarized beams 
is then given by (see. e.g.
\cite{arXiv:0802.0164})
\begin{eqnarray}
\label{rate}
\sigma(beampol)&=&\frac{1-h_{-}}{2}\frac{1+h_{+}}{2}\sigma_{LR}+
\frac{1+h_{-}}{2}\frac{1-h_{+}}{2}\sigma_{RL} \nonumber\\
&=& {\scriptstyle\frac{1}{4}}(1-h_{-}h_{+})\Big( \sigma_{LR}+\sigma_{RL}
+P_{\rm eff}(-\sigma_{LR}+\sigma_{RL})\Big).
\end{eqnarray}
The rate $\sigma(beampol)$ carries an overall helicity alignment factor  
$(1-h_{-}h_{+})$ which drops out when one calculates the normalized 
polarization components of the top quark as in Eqs.(\ref{thresh}) and
(\ref{helimit}). This explains why the polarization depends only
on $P_{\rm eff}$ and not separately on $h_{-}$ and $h_{+}$. Note also that
there is another smaller rate enhancement factor in (\ref{rate}) for negative
values of $P_{\rm eff}$ due to the fact that generally 
$\sigma_{LR}>\sigma_{RL}$.
 
\begin{figure}[ht]
\begin{center}
\includegraphics[scale=1.2]{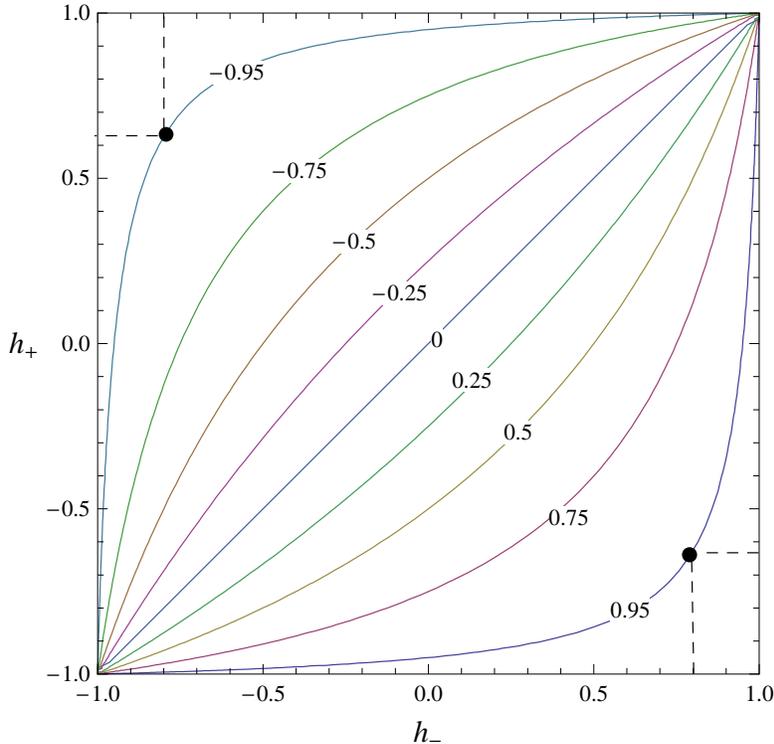} 
\end{center}
\caption{\label{fig:contour}Contour plot in $(h_{+},h_{-})$--plane
}
\end{figure}

Next consider contour plots $P_{\rm eff}={\rm const}$ in the 
$(h_{-},h_{+})$--plane as shown in Fig.\ref{fig:contour}. If one wants large
production rates one has to keep to Quadrants II and IV in 
Fig.\ref{fig:contour} because of the helicity alignment factor 
$(1-h_{-}h_{+})$ in Eq.(\ref{rate}). Fig.\ref{fig:contour} shows 
that near maximal values of 
$P_{\rm eff}$ can be achieved with non-maximal values 
of $(h_{-},h_{+})$. The two examples shown in Fig.\ref{fig:contour} refer
to
\begin{eqnarray}
&&(h_{-}=-0.80,\,h_{+}=+0.625) \qquad {\rm leads \, to}\quad P_{\rm eff}=-0.95
\nonumber\\
&&(h_{-}=+0.80,\,h_{+}=-0.625) \qquad {\rm leads \, to}\quad P_{\rm eff}=+0.95
\end{eqnarray} 
These two options are at the technical limits what can be can achieved 
\cite{arXiv:0905.3066}.
In the next section we shall see that the choice $P_{\rm eff}\sim-0.95$ is
to be preferred since the polarization is more stable against small variations
of 
$P_{\rm eff}$. Furthermore, a negative value of $P_{\rm eff}$ gives yet
another enhancement of the rate~\cite{arXiv:1012.4600} as also indicated in 
the denominator of Eq.(\ref{thresh}) and in the rate formula (\ref{forrate}). 

%%%%%%%%%%%%%%%%%%%%%%%%%%%%%%%%%%%%%%%%%%%%%%%%%%%%%%%%%%%%%%%%%%%%%%%%%%%%%
%%%%%%%%%%%%%%%%%%%%%%%%%%%%%%%%%%%%%%%%%%%%%%%%%%%%%%%%%%%%%%%%%%%%%%%%%%%%%

\section{Stability of polarization against variations of  $P_{\rm eff}$ }
Extrapolations of $|\vec{P}|$ away from $P_{\rm eff}=\pm 1$ are more stable for
$P_{\rm eff}=-1$ than for $P_{\rm eff}=+1$. Take, for example, the magnitude 
of the top quark polarization at threshold Eq.~(\ref{thresh}) and 
differentiate it w.r.t. $P_{\rm eff}$ at
$P_{\rm eff}=\pm 1$. One finds
\begin{equation}
\label{slope}
\frac{d|\vec{P}_{\rm thresh}\,|}{dP_{\rm eff}}=\pm\frac{1\pm A_{LR}}{1\mp A_{LR}}.
\end{equation}
For $P_{\rm eff}=-1$ one has a slope of $-(1-A_{LR})/(1+A_{LR})=-0.42$ while
one has a much larger positive slope of $(1+A_{LR})/(1-A_{LR})=+2.38$ for
$P_{\rm eff}=+1$. This feature persists at higher 
energies~\cite{arXiv:1012.4600}.

%%%%%%%%%%%%%%%%%%%%%%%%%%%%%%%%%%%%%%%%%%%%%%%%%%%%%%%%%%%%%%%%%%%%%%%%%%%%%
\begin{figure}[ht]
\begin{center}
\includegraphics[scale=0.8]{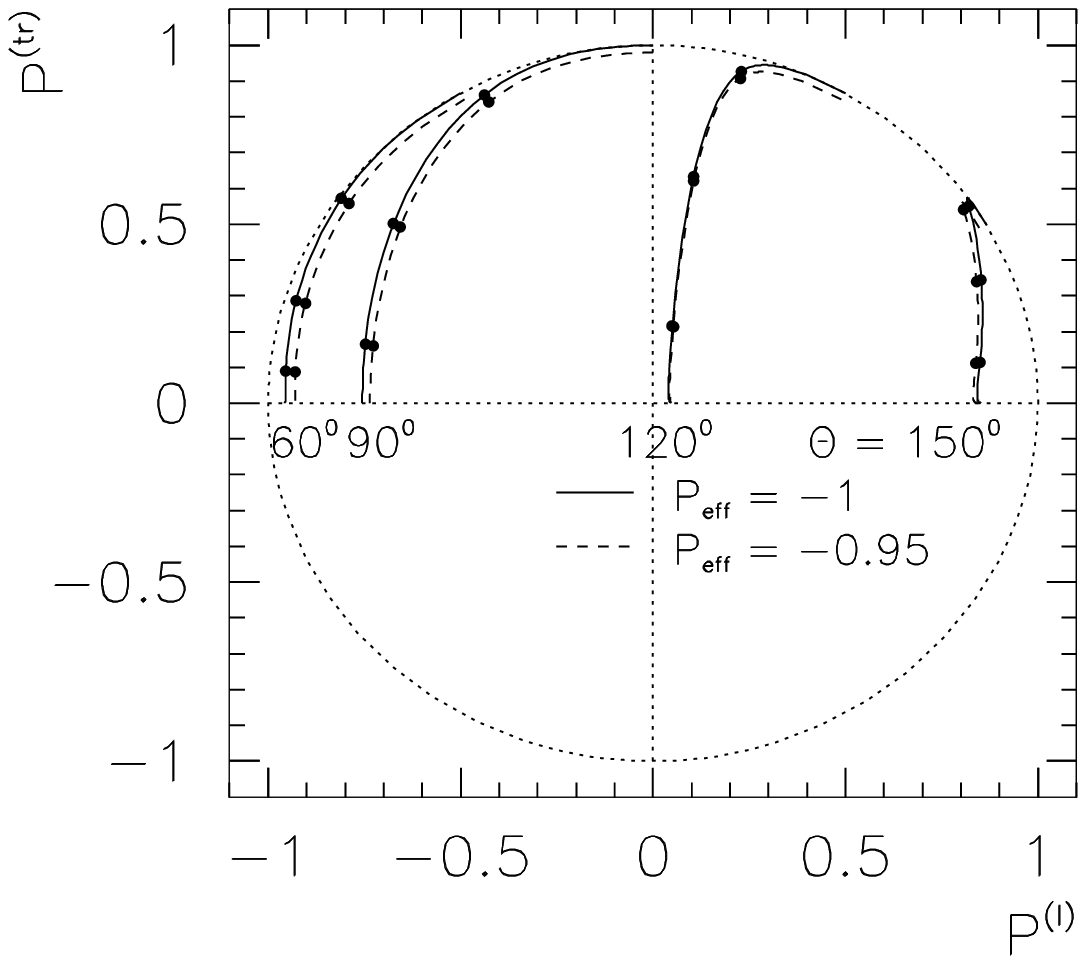} 
\end{center}
\vspace{-1.0cm}
\caption{\label{fig:translong1}Parametric plot of the orientation and the
  length of the
  polarization vector in dependence on the c.m.\ energy $\sqrt{s}$ for values
  $\theta=60^\circ$, $90^\circ$, $120^\circ$, and $150^\circ$ for 
  $P_{\rm eff}=-1$ (solid lines) and $P_{\rm eff}=-0.95$ (dashed lines). The 
  three tics on the trajectories 
  stand for $\sqrt{s}=500\,GeV$, $1000\,GeV$, and $3000\,GeV$.
}
\vspace{1.0cm}
\begin{center}
\hspace{-1.0cm}
\includegraphics[scale=0.8]{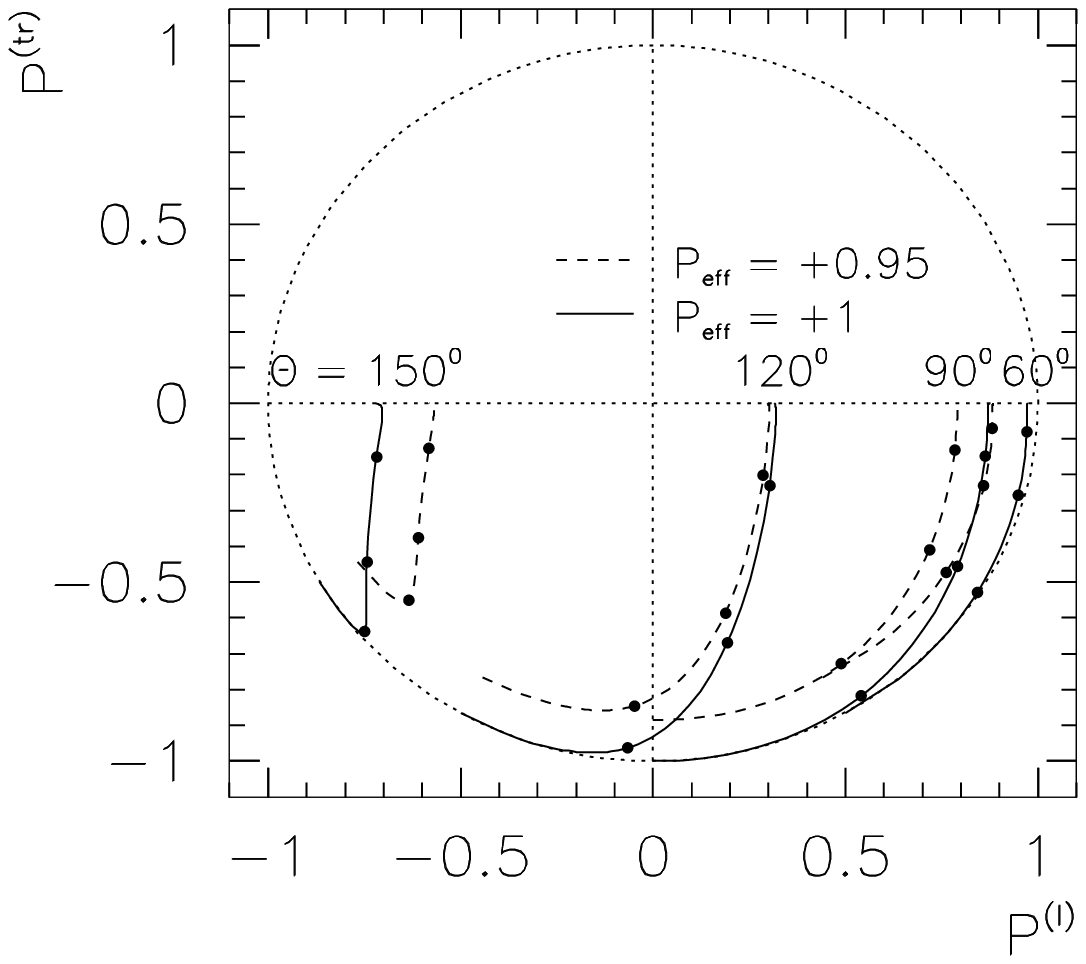} 
\end{center}
\vspace{-1.0cm}
\caption{\label{fig:translong2} Same as Fig.{\ref{fig:translong1}} but for 
  $P_{\rm eff}=+1$ (solid lines) and $P_{\rm eff}=+0.95$ (dashed
  lines). 
}
\end{figure}
%%%%%%%%%%%%%%%%%%%%%%%%%%%%%%%%%%%%%%%%%%%%%%%%%%%%%%%%%%%%%%%%%%%%%%%%%%%%%

\section{Longitudinal and transverse polarization\\
$P^{(\ell)}$ vs. $P^{(tr)}$ for general energies}
In Figs.~\ref{fig:translong1} and \ref{fig:translong2} we plot the longitudinal
component $P^{(\ell)}$ and the transverse component $P^{(tr)}$ of the
top quark polarization for different scattering angles $\theta$ and energies
$\sqrt{s}$ starting from threshold up to the high energy limit.  
$P^{(tr)}$ is the transverse polarization component perpendicular to the
momentum of the top quark in the scattering plane. 
Fig.~\ref{fig:translong1} is drawn for $P_{\rm eff}=(-1,-0.95)$ and 
Fig.~\ref{fig:translong2} for $P_{\rm eff}=(+1,+0.95)$. The apex
of the polarization vector $\vec{P}$ follows a trajectory that starts at
$\vec{P}=P_{\rm thresh}(-\cos\theta,\sin\theta)$ and
$\vec{P}=P_{\rm thresh}(\cos\theta,-\sin\theta)$ for negative and positive
values of $P_{\rm eff}$, respectively, and ends on the line $P^{(tr)}=0$. The 
two $60^{\circ}$ trajectories for
$60^{\circ}$ show that large values of the size of the polarization 
$|\vec{P}|$ close to the maximal value of $1$
can be achieved in the forward region for both $P_{\rm eff}\sim \mp 1$ and
at all energies. However,
the two figures also show that the option $P_{\rm eff}\sim -1$ has to be 
preferred since the $P_{\rm eff}= -1$ polarization is more stable against 
variations of $P_{\rm eff}$ whereas the polarization in 
Fig.~\ref{fig:translong2} has changed considerably
when going from $P_{\rm eff}=1$ to $P_{\rm eff}=0.95$. 

The plots Figs.~\ref{fig:translong1} and \ref{fig:translong2} are drawn for
NLO\, QCD. At NLO there is also a normal component $P^{(n)}$ generated by
the one--loop contribution which, however, is quite small (of $O(3\%))$. 

\section{Summary}
The aim of the investigation in \cite{arXiv:1012.4600} was to maximize and 
to minimize 
the polarization vector of the top quark
$\vec{P}=\vec{P}\,(P_{\rm eff},\sqrt{s},\cos\theta)$
by tuning the beam polarization. Let us summarize our findings.

A. Maximal polarization\\
Large values of $\vec{P}$ can be realized for
$P_{\rm eff}\sim \pm1$ at all intermediate energies. This is particularly true 
in the forward region where the rate is highest. Negative large values for
$P_{\rm eff}$ with aligned beam helicities  
($h_{-}h_{+}$ neg.) are preferred for two reasons.
First there is a further gain in rate apart from the helicity alignment
factor $(1-h_{-}h_{+})$ due to the fact that generally
$\sigma_{LR}>\sigma_{RL}$ as explained after (\ref{rate}) . Second, the 
polarization is more stable against 
variations of $P_{\rm eff}$.

B. Minimal polarization\\
Close to zero values of the polarization vector 
$\vec{P}$ can be achieved for
$P_{\rm eff}\sim 0.4$. Again the forward region is favoured.
In order to maximize the rate for the small polarization choice take quadrant
IV in the $(h_{-},\,h_{+})$--plane.

\vspace*{7.2mm}\noindent
{\bf Acknowledgements:} J.G.K. would like to thank X.~Artru and E.~Christova
for discussions and G.~Moortgat-Pick for encouragement. We thank
B.~Meli\'c and S.~Prelovsek for their participation in this project. 
The work of S.~G. is supported by the Estonian
target financed project No.~0180056s09, by the Estonian Science Foundation
under grant No.~8769 and by the Deutsche Forschungsgemeinschaft (DFG)
under grant 436 EST 17/1/06.

\end{document}